\documentclass[aps,amsmath,amssymb,reprint]{revtex4-1}

\usepackage{graphicx}
\usepackage{dcolumn}
\usepackage{bm}

\usepackage{mathtools}

\usepackage[utf8]{inputenc}
\usepackage[T1]{fontenc}
\usepackage{mathptmx}
\usepackage{xcolor}
\definecolor{color1}{HTML}{1f77b4} 
\definecolor{color2}{HTML}{FF7F0E} 
\definecolor{color3}{HTML}{3ae03a} 
\definecolor{color4}{HTML}{D62728} 
\definecolor{color5}{HTML}{9467BD} 
\usepackage{siunitx}
\usepackage{color}

\usepackage{tikz}
\usetikzlibrary{shapes,arrows}
\usetikzlibrary{spy,decorations.fractals,shapes.misc}
\usetikzlibrary{decorations.markings}
\usepackage{pgfplots}
\usepgfplotslibrary{groupplots}
\usepgfplotslibrary{colormaps}
\usepgfplotslibrary{colorbrewer}
\pgfplotsset{cycle list/Dark2}
\pgfplotsset{compat=newest,/pgf/number format/.cd,1000 sep={}}
\pgfplotsset{filter discard warning=false}
\usepackage[caption=false]{subfig}
\usepgfplotslibrary{colormaps}
\usepgfplotslibrary{colorbrewer}
\pgfplotsset{cycle list/Dark2}
\usepackage[capitalise]{cleveref}
\usetikzlibrary{matrix}

\begin{document}

\title[Crank-Nicolson BGK Integrator]{Crank-Nicolson BGK Integrator for Multi-Scale Particle-Based Kinetic Simulations} 

\author{M. Pfeiffer}
\thanks{Corresponding author.}
 \email{mpfeiffer@irs.uni-stuttgart.de}
 \affiliation{%
   Institute of Space Systems, University of Stuttgart, Pfaffenwaldring 29, 70569
   Stuttgart, Germany
 }%

\author{F. Garmirian}
 \email{garmirianf@irs.uni-stuttgart.de}
 \affiliation{%
   Institute of Space Systems, University of Stuttgart, Pfaffenwaldring 29, 70569
   Stuttgart, Germany
 }%

\author{T. Ott}
 \email{ottt@irs.uni-stuttgart.de}
 \affiliation{%
   Institute of Space Systems, University of Stuttgart, Pfaffenwaldring 29, 70569
   Stuttgart, Germany
 }%

\date{\today}
\begin{abstract}
Solving the Bhatnagar-Gross-Krook (BGK) equation with a stochastic particle approach enables efficient and flexible simulations of flows in the transition regime, between continuum and free molecular flow. However, the usual first-order operator splitting between particle movement and relaxation imposes restrictions on the time step, causing the computational cost to increase with the gas density. The Crank-Nicolson stochastic particle BGK (CN-SPBGK) method is introduced here as an advanced particle-based kinetic solver designed for multi-scale gas flow simulations. This method integrates the BGK equation with second-order accuracy across all Knudsen number regimes without requiring additional parameters, while asymptotically preserving the Navier-Stokes flux in the continuum regime. Comparisons with pre-existing particle BGK methods are conducted on several test cases, with CN-SPBGK demonstrating more consistent convergence and accuracy.
\end{abstract}

\maketitle
\section{Introduction}
The Direct Simulation Monte Carlo (DSMC) method has long been the benchmark for simulating rarefied, non-equilibrium gas flows \cite{Bird1994}. The core of DSMC lies in using discrete particle collisions to statistically replicate molecular interactions, offering a robust particle-based approach to phase-space treatment. This enables straightforward incorporation of internal energy modes, chemical reactions, and complex boundary interactions. While DSMC has proven effective across a wide range of rarefaction regimes, it encounters challenges near the continuum limit, where resolving short mean free paths and high collision frequencies becomes computationally intensive. Consequently, flows with a broad Knudsen number range, involving both continuum and rarefied phenomena, remain difficult to simulate efficiently with DSMC alone.

To address this limitation, various methods have been developed. One direct approach is hybrid coupling between DSMC and continuum solvers based on the Navier-Stokes equations \cite{hash1995hybrid,carlson2004hybrid, tatsios2025dsmc, VASILEIADIS2024103669}. However, such coupling is complex: Navier-Stokes solvers use macroscopic variables, while DSMC relies on particle-based probabilities, leading to interface challenges that can limit the general applicability and robustness of these methods.

Alternatively, multi-scale methods based on phase-space discretization, such as Discrete Velocity Methods (DVMs) \cite{mieussens2000discrete}, have shown promise. Recently developed methods like the Unified Gas-Kinetic Scheme (UGKS) \cite{xu2010unified,chen2015comparative,zhu2021first} and the Discrete Unified Gas-Kinetic Scheme (DUGKS) \cite{guo2013discrete,guo2021progress} efficiently approximate the collision term through Bhatnagar-Gross-Krook (BGK) models. By providing deterministic, noise-free solutions, these methods are well-suited for low-Mach flows, though they may become computationally demanding for high-Mach non-equilibrium flows requiring large velocity-space domains.

The Unified Gas-Kinetic Wave-Particle (UGKWP) method \cite{liu2020unified,chen2020three} offers a hybrid solution, representing the non-equilibrium portion of the velocity distribution with particles, thereby reducing the computational cost associated with velocity discretization. This approach enables noise-free hydrodynamic solutions in equilibrium regions by limiting particle representation to non-equilibrium zones. However, combining particle and DVM treatments presents implementation challenges, especially when adapting the approach for established DSMC solvers.

On the other side, particle-based BGK \cite{Pfeiffer2018,zhang2019particle,pfeiffer2019evaluation} and Fokker-Planck (FP) methods \cite{gorji2014efficient,gorji2021entropic,mathiaud2016fokker} have garnered attention as they integrate easily with DSMC and show efficiency in moderate- to low-Knudsen number flows. Notable BGK models include particle-based approaches such as the Ellipsoidal Statistical BGK (ES-BGK) \cite{gallis2011investigation,gallis2000application,burt2006evaluation} and Shakhov BGK methods \cite{Pfeiffer2018,fei2020benchmark}. In FP methods, significant developments have been made in the Entropic Fokker-Planck (EFP) model \cite{gorji2021entropic}, the cubic FP model \cite{gorji2011fokker}, the Ellipsoidal Statistical FP (ESFP) model \cite{mathiaud2016fokker}, and Bogomolov’s model \cite{bogomolov2009fokker}. Despite reducing the computational expense associated with collisions in the continuum regime, these methods still require fine resolution due to the typical first-order particle evolution resulting from separate free-flight and collision/relaxation sub-steps. On the Fokker-Planck side, methods have recently been developed which are second order and asymptotically preserving with respect to the Navier-Stokes equations at small Knudsen numbers. These thus allow much coarser time and space discretisations~\cite{kim2024second,cui2025multiscale}.
In addition, direct extensions of the DSMC method in the form of the time-relaxed Monte Carlo method~\cite{pareschi2001time} have been introduced recently, which show direct asymptotic behavior with respect to the Navier-Stokes equations and thus may be a good way to solve multiscale problems in the future~\cite{FEI2023112128,fei2024navier}.

This paper focuses on BGK-based particle methods. One BGK-based method that has proven particularly efficient in the low Knudsen number regime is the Unified Stochastic Particle (USP) method~\cite{fei2020unified,fei2021,feng2023spartacus}. This method introduces an additional collision operator during particle movement, enabling correction of operator splitting between advection and relaxation in the continuum regime, and allowing for significantly larger time steps. The method has been extended to inner degrees of freedom~\cite{fei2022unified}, higher order methods~\cite{fei2022high} and coupled with DSMC~\cite{fei2021hybrid}. As discussed in Section~\ref{sec:secordpart}, an additional parameter is required to control the transition between low and high Knudsen numbers for this method.
Another approach for particle methods is Exponential Differencing (ED-SPBGK), which enables implicit handling of the stiff collision term while ensuring positivity preservation of the method and all distribution functions~\cite{pfeiffer2022exponential}. However, due to the limitations in influencing flux for particles (compared to DVM~\cite{garmirian2023asymptotic}), the ED-SPBGK variant is less efficient at small Knudsen numbers than ED-DVM or USP.

\section{Theory}
The Boltzmann equation models the behavior of a monatomic gas flow by describing its distribution function $f = f(\mathbf{x}, \mathbf{v}, t)$, which depends on position $\mathbf{x}$, velocity $\mathbf{v}$, and time $t$. The equation is given by

\begin{equation}
\frac{\partial f}{\partial t} + \mathbf{v} \cdot \frac{\partial f}{\partial \mathbf{x}} = \left.\frac{\delta f}{\delta t}\right|_\mathrm{coll},
\end{equation}

where external forces are neglected. The term $\left.\delta f / \delta t\right|_\mathrm{coll}$ represents the Boltzmann collision integral:

\begin{equation}
\left.\frac{\partial f}{\partial t}\right|_{\mathrm{coll}} = \int_{\mathbb{R}^3} \int_{S^2} B \left[f(\mathbf{v}') f(\mathbf{v}_*') - f(\mathbf{v}) f(\mathbf{v}_*)\right] \, d\mathbf{n} \, d\mathbf{v}_*.
\end{equation}

Here, $S^2 \subset \mathbb{R}^3$ denotes the unit sphere, $\mathbf{n}$ is a unit vector representing the direction of scattered velocities, $B$ is the collision kernel, and the superscript $'$ indicates post-collision velocities. The nonlinearity of this collision term and the high dimensionality of the solution domain contribute to the computational complexity of the Boltzmann collision integral. To address this, the Direct Simulation Monte Carlo (DSMC) method simplifies the collision integral by approximating it through Monte Carlo sampling of collisions between randomly selected particles that represent the gas flow.

\subsection{BGK Approximation}

The BGK model approximates the collision term using a relaxation form, in which the distribution function $f$ relaxes toward a target distribution $f^t$:

\begin{equation}
\left.\frac{\partial f}{\partial t}\right|_{\text{Coll}} = \Omega = \nu \left(f^t - f\right)= \frac{1}{\tau} \left(f^t - f\right)
\label{eq:bgkmain}
\end{equation}

where $\nu$ is the relaxation frequency or $\tau$ the relaxation time. In the original BGK model, the target distribution is assumed to be Maxwellian:

\begin{equation}
f^M = n \left(\frac{m}{2 \pi k_B T}\right)^{3/2} \exp\left[-\frac{m \mathbf{c} \cdot \mathbf{c}}{2 k_B T}\right],
\label{eq:maxwelldist}
\end{equation}

with number density $n$, molecular mass $m$, temperature $T$, Boltzmann constant $k_B$, and thermal velocity $\mathbf{c} = \mathbf{v} - \mathbf{u}$, where $\mathbf{u}$ is the average flow velocity \cite{bhatnagar1954model}. The relaxation frequency $\nu$ is related to the viscosity by

\begin{equation}
\mu = \frac{n k_B T}{\nu}.
\end{equation}

Using the Maxwellian distribution as the target results in a fixed Prandtl number $\textrm{Pr} = \mu c_P / K = 1$, whereas the Prandtl number for monatomic gases is approximately $2/3$ \cite{vincenti1965introduction}. To address this discrepancy, several extensions of the BGK model have been proposed. Some of these models adjust the target distribution function, such as the ellipsoidal statistical BGK model \cite{holway1966new} or the Shakhov BGK model \cite{shakhov1968generalization}, while others modify the relaxation frequency, making it a function of microscopic velocities as described by \citet{struchtrup1997bgk}.

The different BGK models with correct Prandtl number are built using higher moments of the distribution function, namely the traceless pressure tensor $p_{\langle ij \rangle}$ and the heatflux $\mathbf q$, with
\begin{equation}
\mathbf q = \frac{1}{2}m\int \mathbf c (\mathbf c \cdot \mathbf c) f\,d\mathbf c    
\end{equation}
and
\begin{equation}
p_{\langle ij \rangle}=m\int c_{\langle i}c_{j\rangle}f\,d\mathbf c.
\end{equation}

\subsection{Particle based BGK solver}
Although the BGK relaxation model has a much simpler structure compared to the Boltzmann collision integral, stiff relaxation can still arise when explicit time integrations are used. A major challenge in solving the BGK equation is that the source term becomes stiff as the relaxation time $\tau=1/\nu$ decreases, which requires resolution through explicit time integration.

Moreover, the equation should be integrated using an asymptotically preserving second-order method to ensure that the method captures the correct gas viscosity for time steps larger than the relaxation time, i.e., to converge asymptotically to the Navier-Stokes equations.

In the stochastic particle BGK method (SPBGK), the time integration of Eq.~\eqref{eq
} is typically carried out by assuming a constant target distribution\cite{gallis2011investigation,gallis2000application}: \begin{equation} f(\mathbf v, \mathbf x, t+\Delta t) = e^{-\nu\Delta t}f(\mathbf v, \mathbf x, t) + (1-e^{-\nu\Delta t})f^t(\mathbf v, \mathbf x, t). \label{eq
} \end{equation} The advection step employs a first-order operator-splitting approach, meaning that the particles of $f(\mathbf v, \mathbf x, t+\Delta t)$ are advected along their trajectories for the entire time step $\Delta t$ to yield $f(\mathbf v, \mathbf x+\mathbf v \Delta t, t+\Delta t)$.

This first-order time integration offers some advantages in particle methods. For $\nu\Delta t \ll 1$, the time integration Eq.\eqref{eq
} reduces to the forward Euler method: \begin{equation} f(\mathbf v, \mathbf x, t+\Delta t) = (1-\nu\Delta t)f(\mathbf v, \mathbf x, t) + \nu\Delta t f^t(\mathbf v, \mathbf x, t). \label{eq
} \end{equation} However, unlike the forward Euler method in Eq.\eqref{eq
}, the prefactors of $f(\mathbf v, \mathbf x, t)$ and $f^t$ in Eq.\eqref{eq
} remain positive even when $\nu\Delta t > 1$. In the context of stochastic particle methods, Eq.\eqref{eq
} is straightforward to implement: each particle in a cell acquires a new velocity sampled from $f^t$ with a well-defined probability of $(1-e^{-\nu\Delta t})$.

\subsubsection{Second Order Particle BGK schemes}
\label{sec:secordpart}
A particle-based BGK variant that performs highly efficiently in the continuum regime is the Unified Stochastic Particle (USP) method~\cite{fei2020unified,fei2021,feng2023spartacus}. In USP, an additional collision term is introduced, where the current distribution function for the flux is approximated by a Grad-13 expansion. This approach virtually enables simultaneous treatment of the advection and relaxation processes. The selection of the Grad-13 distribution also ensures asymptotic convergence to the Navier-Stokes limit.

This additional collision term is designed to satisfy the Navier-Stokes equations within the continuum domain using a second-order time integration, while the conventional first-order SPBGK method is applied in the rarefied domain. Furthermore, a multi-scale parameter is defined to allow a smooth transition between the standard SPBGK method and the Grad-13 approximation.

In the transition regime, depending on the choice of the multi-scale parameter, either the method fully reverts to SPBGK, achieving only first-order accuracy, or it results in a blend of the SPBGK method and the additional collision term. The error due to interpolation between the standard SPBGK method and the added collision term remains somewhat ambiguous. Nevertheless, the performance of the USP method in continuum regions is notably effective, as both the collision operator and the flux term are structured to match the Navier-Stokes flux accurately.

Another second-order particle BGK method is the Exponential Differencing (ED-SPBGK) method~\cite{pfeiffer2022exponential}. This method benefits from exponential time integration, making it a second-order scheme that is inherently positive, which is well-suited for particle simulations. The ED-SPBGK method maintains second-order accuracy for the collision term without flux even at high Knudsen numbers and does not require additional parameters for transitioning between high and low Knudsen number regimes.

However, in low Knudsen number regions, the ED-SPBGK method is not as efficient as the USP method. While the collision operator integration is second-order implicit, the flux term, unlike in the USP method, does not revert to the Navier-Stokes limit at small Knudsen numbers.

While in the Discrete Velocity Method (DVM) it is possible to construct the flux between cells using the ED-SPBGK variant in a way that is asymptotic-preserving to the Navier-Stokes equations~\cite{garmirian2023asymptotic}, making the ED-SPBGK approach very efficient, in particle methods, the flux is determined by the particle motion. This limits the flexibility for manipulation, making it difficult to maintain the Navier-Stokes flux at low Knudsen numbers, as is achievable in the DVM approach or the USP method.
The goal, therefore, is to develop a particle-based method that achieves second-order accuracy across all Knudsen number regimes without requiring additional parameters, as in the ED-SPBGK particle variant. At the same time, it should replicate the Navier-Stokes flux, as the USP method does, to ensure high efficiency at low Knudsen numbers.

\subsubsection{CN-SPBGK}
The CN-SPBGK method possesses the desired properties: it is second-order accurate across all Knudsen number regimes, requires no additional parameters, and converges to the Navier-Stokes flux at low Knudsen numbers, similar to the USP method. The entire approach is based on UGKS and DUGKS\cite{xu2010unified,guo2013discrete}.

The Crank-Nicolson method is employed for time integration as follows: 
\begin{eqnarray} &f(\mathbf v, \mathbf x+\mathbf v \Delta t, t+\Delta t) = f(\mathbf v, \mathbf x, t) \nonumber
\\ & + \frac{\Delta t}{2}\left(\Omega(\mathbf v, \mathbf x+\mathbf v \Delta t, t+\Delta t) + \Omega(\mathbf v, \mathbf x, t) \right). \label{eq:CN} 
\end{eqnarray}

Next, let us introduce two additional distributions: \begin{eqnarray} \tilde{f} &=& f - \frac{\Delta t}{2}\Omega = \frac{2\tau+\Delta t}{2\tau}f-\frac{\Delta t}{2\tau}f^t\ \textrm{and} \label{eq:aux1} \\\ 
\hat{f} &=& f + \frac{\Delta t}{2}\Omega = \frac{2\tau-\Delta t}{2\tau}f+\frac{\Delta t}{2\tau}f^t, \label{eq:aux2} \end{eqnarray} 
where $\tau=1/\nu$ is the relaxation time.

Substituting the distributions defined above into Eq.~\eqref{eq:CN}, it becomes evident that $\tilde{f}(\mathbf v, \mathbf x+\mathbf v \Delta t, t+\Delta t)$ corresponds to particles of $\hat{f}(\mathbf v, \mathbf x, t)$ that are advected along their trajectories for a time step $\Delta t$, i.e., \begin{equation} \tilde{f}(\mathbf v, \mathbf x+\mathbf v \Delta t, t+\Delta t)=\hat{f}(\mathbf v, \mathbf x, t) . \end{equation}

This approach provides an implicit integration of the coupled advection and relaxation, theoretically making it suitable for stochastic particle methods as well.
Here, $\hat f$ is essentially equivalent to $f^{n+1/2}$, representing the collision operator applied with a half time step. When the particles are then advected to obtain $\tilde f$, this effectively constructs a classic leapfrog scheme.
Since the asymptotic preserving of the approach for the collision term has already been discussed in detail in~\cite{guo2013discrete}, only the different flow term in contrast to the non-particle-based DUGKS and UGKS needs to be considered here. However, as shown in Appendix~\ref{sec:appflux}, the natural leapfrog behavior resulting from the CN approach, as in the USP method, results in the correct Navier-Stokes flux for small Knudsen numbers. As in the original DUGKS and UGKS variants, \eqref{eq:aux1} and \eqref{eq:aux2} can be combined into one step resulting in:
\begin{equation}
    \hat f = \frac{2\tau-\Delta t}{2\tau + \Delta t}\tilde f + \frac{2 \Delta t}{2\tau  + \Delta t} f^t.
    \label{eq:fullCN}
\end{equation}
Now the distributions $\hat f$ and $\tilde f$ can be tracked directly and only one relaxation step per time step is necessary. However, the moments of the target function $f^t$ are dependent on $f$. While density, mean flow velocity $\mathbf{u}$ and temperature $T$ are identical for $\hat f$, $\tilde f$ and $f$, the higher moments such as pressure tensor $p_{ij}$ and heat flux $\mathbf{q}$ require conversions depending on the selected target function as discussed in~\cite{guo2013discrete} and the following section~\ref{sec:impl}.

\section{Implementation}
\label{sec:impl}
Up to this point, we have followed the DUGKS approach, demonstrating that the resulting flux, due to the half time step applied before particle movement, produces the desired flux. However, a significant challenge now arises in implementing the particle aspect of Eq.~\eqref{eq:fullCN}. Here, a distinction must be made: as long as $2\tau > \Delta t$, the standard approach of BGK particle methods can be applied. This involves interpreting the prefactors as probabilities. Specifically, a random number $P \in (0,1]$ is drawn and compared with the factor $\frac{2\Delta t}{2\tau + \Delta t}$ from Eq.~\eqref{eq:fullCN}. If $P$ is smaller, the particle is re-sampled according to the target function $f^t$; otherwise, it remains unchanged.

However, this concept breaks down when $2\tau < \Delta t$, as this results in negative prefactors.
In this case, $\hat f$ is directly approximated by a Grad-13 distribution, and all particles are sampled from the Grad-13 distribution with the appropriately determined pressure tensor $\hat{p}_{ij}$ and heat flux $\hat{\mathbf{q}}$
\begin{equation}
    f^{G13}(\mathbf{v}) =f^M\left[1+\frac{ c_i c_j \hat p_{\langle ij\rangle}}{2\rho (RT)^2} + \frac{\mathbf c \cdot \hat{\mathbf q}}{5\rho (RT)^2}\left(\frac{\mathbf c^2}{RT}-5\right)\right].
    \label{eq:grad13}
\end{equation}

This approximation assumes that higher moments beyond the heat flux can be neglected in the resulting distribution function. However, since this approach is applied only when $\Delta t > 2\tau$, it is used exclusively for very large time steps or very small relaxation times, making this assumption reasonable.
The two most common target distributions are the ellipsoidal
statistical BGK (ESBGK) and the Shakhov model (SBGK). In the following, the corresponding calculations to get ${p}_{ij}$ and heat flux ${\mathbf{q}}$ from the f as well as $\hat{p}_{ij}$ and heat flux $\hat{\mathbf{q}}$ for ESBGK and SBGK are discussed. How exactly particles can be sampled from the Grad-13 function and the SBGK or ESBGK target function is described in various ways in~\cite{Pfeiffer2018}.

\subsection{ESBGK}
The ES-BGK distribution is given as~\cite{holway1966new} 
\begin{equation}
f^{ES}=\frac{n}{\sqrt{\det (2\pi\lambda_{ij})}}\exp\left[-\frac{1}{2}\lambda_{ij}^{-1}c_ic_j\right]
\label{eq:esbgkdist}
\end{equation}  
with the matrix 
\begin{equation}
\lambda_{ij} = \frac{k_BT}{m}\delta_{ij} + \left(1-\frac{1}{\textrm{Pr}}\right)\frac{p_{\langle ij\rangle}}{\rho}.
\end{equation}
Here, $\delta_{ij}$ is the Kronecker delta, $\textrm{Pr}$ is the Prandtl number and $\rho$ is the mass density. Furthermore, the relaxation time is given as
\begin{equation}
    \tau = \frac{\mu}{n k_B T Pr}
\end{equation}
In the case of $\Delta t < 2\tau$, only the pressure tensor is required, as only this is included in the target distribution in the ESBGK model. Equation \eqref{eq:aux1} is used to calculate ${p}_{ij}$:
\begin{equation}
    f=\frac{2\tau}{2\tau+\Delta t} \tilde f + \frac{\Delta t}{2\tau+\Delta t} f^t
\end{equation}
Using this equation as well as \eqref{eq:esbgkdist} and the integration to obtain ${p}_{ij}$ results in
\begin{eqnarray}
    p_{ij}&=&\frac{2\tau}{2\tau+\Delta t} \tilde p_{ij} + \frac{\Delta t}{2\tau+\Delta t}\left(\frac{Pr-1}{Pr}p_{ij}+\frac{1}{Pr}\frac{k_BT}{m\rho}\delta_{ij}\right) \\
    \Rightarrow p_{ij} &=& \frac{2\tau Pr}{2\tau Pr + \Delta t}\tilde p_{ij}+ \frac{\Delta t}{2\tau Pr + \Delta t} \frac{k_BT}{m\rho}\delta_{ij}
    \label{eq:esbgksmalldt}
\end{eqnarray}
This equation can be used to calculate the $p_{ij}$ required for the target function directly from the tracked $\tilde f$ and the $\tilde p_{ij}$ sampled from it.

For case $\Delta t > 2\tau$, the resulting heat flux $\hat{\mathbf{q}}$ and pressure tensor $\hat p_{ij}$ are now calculated directly from the given $\tilde f$ for the Grad13 approximation of $\hat f$~\eqref{eq:grad13} using \eqref{eq:fullCN} and \eqref{eq:esbgksmalldt}:
\begin{eqnarray}
    \hat{\mathbf{q}} &=& \frac{2\tau-\Delta t}{2\tau + \Delta t} \tilde{\mathbf{q}}\label{eq:momentsesbgkq}\\
    \hat p_{\langle ij \rangle} &=& \frac{2\tau-\Delta t}{2\tau + \Delta t} \tilde p_{\langle ij \rangle} + \frac{2\Delta t}{2\tau+\Delta t}\frac{Pr-1}{Pr}\frac{2\tau Pr}{2\tau Pr + \Delta t}\tilde p_{\langle ij \rangle}
    \label{eq:momentsesbgkp}
\end{eqnarray}
Thus, a corresponding particle can be sampled directly from the Grad-13 approximation of $\hat f$ according to \eqref{eq:grad13} with the moments of \eqref{eq:momentsesbgkq} and \eqref{eq:momentsesbgkp}.

\subsection{SBGK}
In the case of the S-BGK model~\cite{shakhov1968generalization}, the target distribution is defined as
\begin{equation}
f^S=f^M\left[1+(1-\textrm{Pr})\frac{\mathbf c \cdot \mathbf q}{5 \rho (RT)^2}\left(\frac{\mathbf c \cdot \mathbf c}{RT}-5\right)\right]
\end{equation} 
with the relaxation time being:
\begin{equation}
    \tau = \frac{\mu}{n k_B T}.
\end{equation}
In the SBGK case, for $\Delta t < 2\tau$, only the heat flux $\mathbf{q}$ is required, which, as already described in~\cite{guo2013discrete}, can be easily calculated from $\mathbf{\tilde{q}}$ using the same procedure as in the ESBGK case:
\begin{eqnarray}
    \mathbf{q}&=& \frac{2\tau}{2\tau+\Delta t} \mathbf{\tilde q} + (1-Pr)\frac{\Delta t}{2\tau+\Delta t} \mathbf{q}\\
    \Rightarrow \mathbf{q} &=& \frac{2\tau}{2\tau+\Delta t Pr} \mathbf{\tilde q}
    \label{eq:sbgksmalldt}
\end{eqnarray}
In the case of $\Delta t > 2\tau$, the following moments result for the SBGK model for the Grad-13 approximation of $\hat f$:
\begin{eqnarray}
    \mathbf{\hat q} &=& \frac{2\tau-\Delta t}{2\tau + \Delta t} \mathbf{\tilde q}+(1-Pr)\frac{2\Delta t}{2\tau+\Delta t}\frac{2 \tau}{2\tau + \Delta t Pr} \mathbf{\tilde q}\label{eq:momentssbgkq}\\
    \hat p_{\langle ij \rangle} &=& \frac{2\tau-\Delta t}{2\tau + \Delta t} \tilde p_{\langle ij \rangle}.
    \label{eq:momentssbgkp}
\end{eqnarray}

\subsection{Typical time step and conservation of energy}

Details of the sampling process, along with an in-depth discussion of possible energy and momentum conservation schemes for the particle BGK method, can be found in \citet{Pfeiffer2018}.
To ensure momentum and energy conservation, the final velocities of the particles are adjusted according to

\begin{equation}
\mathbf{v}_i^* = \mathbf{u} + \alpha(\mathbf{v}'_i - \mathbf{u}'),
\label{eq:transenmom}
\end{equation}

where $\mathbf{u} = \sum_{i=1}^N \mathbf{v}_i / N$ represents the bulk flow velocity before relaxation, $\mathbf{v}'_i$ is the particle velocity after relaxation (uncorrected), and $\mathbf{u}' = \sum_{i=1}^N \mathbf{v}'_i / N$. Note that $\mathbf{v}'_i = \mathbf{v}_i$ if no relaxation occurs for particle $i$. Given that

\begin{equation}
\sum_{i=1}^N (\mathbf{v}'_i - \mathbf{u}') = 0,
\end{equation}

Equation \eqref{eq:transenmom} guarantees momentum conservation. Energy conservation is ensured by selecting $\alpha$ as

\begin{equation}
\alpha = \sqrt{\frac{T}{T'}},
\end{equation}

where $T$ denotes the temperature before relaxation and $T'$ is the temperature after the relaxation process (in the absence of energy correction).

To achieve second-order accuracy in space, the sampled macroscopic data — such as density, flow velocity, temperature, pressure tensor, and heat flux — are linearly interpolated. Various interpolation methods are possible; however, in this work, the stochastic interpolation from~\cite{fei2020unified, fei2021, feng2023spartacus} was used. For the stochastic interpolation, the macroscopic values are sampled per cell as in the standard SPBGK method. Then, particles are moved a random distance between 0 and half of a cell width in a specific direction, and the values of the cell at the resulting location are assigned as the particle’s properties. This method allows for simple implementation and easy handling of grid adaptations such as octrees or sub-cells. The time step of the CN-SPBGK method proceeds as follows:

\begin{enumerate}
    \item The macroscopic quantities are sampled and stored per cell from the particles.
    \item Stochastic interpolation is applied to the particles to define macroscopic quantities at the particle locations.
    \item Based on the macroscopic values, the relaxation time for each particle is calculated according to either the SBGK or ESBGK model.
    \item Depending on the time step and relaxation time:
    \begin{enumerate}
        \item If $\Delta t < 2\tau$, a random number is drawn per particle. If the random number is less than $2\Delta t / (2\tau + \Delta t)$, the particle undergoes relaxation; otherwise, it remains unchanged. If it relaxes, it relaxes according to the standard target functions of ESBGK or SBGK, using the scaled pressure tensor \eqref{eq:esbgksmalldt} or heat flux \eqref{eq:sbgksmalldt}.
        \item If $\Delta t > 2\tau$, the particle is directly sampled from the Grad-13 distribution, with the corresponding scaled pressure tensor and heat flux for ESBGK \eqref{eq:momentsesbgkq} and \eqref{eq:momentsesbgkp} or SBGK \eqref{eq:momentssbgkq} and \eqref{eq:momentssbgkp}.
    \end{enumerate}
    \item Energy and momentum conservation of the particles as described.
    \item Particles move and interact with the boundaries as usual in the DSMC method.
\end{enumerate}

\section{Results}
\subsection{Couette flow}
The performance of the CN-SPBGK method is compared with that of other methods using a 1D Couette flow test case.
In this test case, two parallel plates, separated by a distance of $L = \SI{1}{\meter}$ in the $x$-direction, move in opposite directions with velocities $v_y = \SI{500}{\meter\per\second}$ and $v_y = -\SI{500}{\meter\per\second}$, respectively in the y-direction.
The plates have a constant temperature of $T_w = \SI{273}{\kelvin}$ and are modeled as diffuse boundaries.
The simulations are performed using Argon gas, with properties given in Table \ref{tab:argon}.

Two flow regimes are simulated: a rarefied regime with a high Knudsen number $Kn = 0.1$ (case 1), and a continuum regime with $Kn = 0.001$ (case 2).
A summary of the conditions for these test cases can be found in Table \ref{tab:couettecond}.
The CN-SPBGK method, utilizing either Shakhov or ES-BGK as the target distribution functions, is compared against the standard SPBGK method, the USPBGK method, and the exponential differencing ES-BGK method (ED-SPBGK).
It should be noted that, in contrast to the original paper, the USPBGK method was used without the Knudsen-number-based switching between USPBGK in dense regimes and SPBGK in rarefied regimes.

To exclusively examine the impact of time integration on the performance of the various methods, identical spatial interpolation was used across all simulations.

\begin{table}
\caption{Start conditions of Couette flow simulations.}\label{tab:couettecond}
  \begin{ruledtabular}\renewcommand*{\arraystretch}{1.4}
    \begin{tabular}{ccccc}
        &  \# computational cells  & number density $n_0\,/\, \SI{}{\per\meter\cubed}$ & Kn \\
      \hline
      Case 1 & 20 & $1.37\cdot10^{19}$ & 0.1 \\
      Case 2 & 50 & $1.37\cdot10^{21}$ & 0.001 \\
    \end{tabular}
  \end{ruledtabular}
\end{table}

\begin{table}
\caption{Species-specific VHS parameters used for the simulation.}\label{tab:argon}
  \begin{ruledtabular}\renewcommand*{\arraystretch}{1.4}
    \begin{tabular}{cccc}
      Species & $d_{VHS} \,/\, \SI{}{\meter}$ & $T_{VHS} \,/\, \SI{}{\kelvin}$ & $\omega_{VHS}$ \\
    Ar & $\SI{4.05E-10}{}$ & $273$ & $0.77$ \\
    \end{tabular}
  \end{ruledtabular}
\end{table}

\subsubsection{Rarefied case ($Kn=0.1$)}
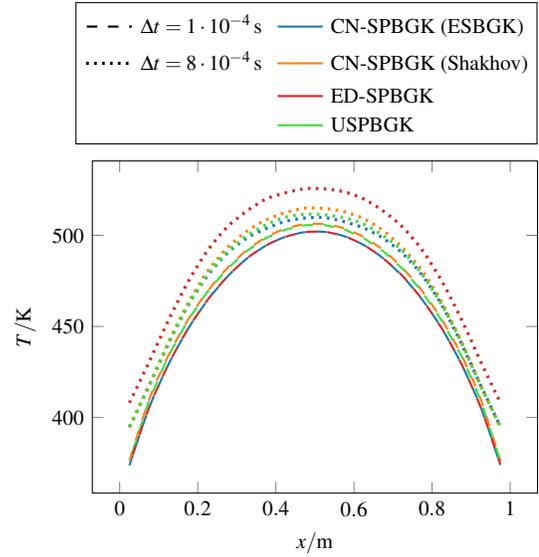
\begin{figure}
	\centering
	\begin{tikzpicture}
  \begin{axis}[
  scaled ticks=false,
  width=7.5cm,
  height=6cm,
  xlabel={$x / \text{m}$},
  xticklabel style={
    /pgf/number format/fixed,
    /pgf/number format/precision=2
  },
  ylabel={$T / \text{K}$},
  xmin=-0.07,
  xmax=1.07,
  cycle list name=exotic,
  font=\footnotesize,
  ]

  \addplot[mark=none,dash pattern= on 1pt off 2pt,smooth,color=color1,very thick] table [col sep=comma, x=x, y expr=\thisrow{T}/1] {data/Couette-01-8e-4-CN.csv}; \label{plot:01-CN-1}
  \addplot[mark=none,dash pattern= on 4pt off 4pt,smooth,color=color1,thick] table [col sep=comma, x=x, y expr=\thisrow{T}/1] {data/Couette-01-1e-4-CN.csv}; \label{plot:01-CN-2}

  \addplot[mark=none,dash pattern= on 1pt off 2pt,smooth,color=color2,very thick] table [col sep=comma, x=x, y expr=\thisrow{T}/1] {data/Couette-01-8e-4-CN-S.csv}; \label{plot:01-CN-S-1}
  \addplot[mark=none,dash pattern= on 4pt off 4pt,smooth,color=color2,thick] table [col sep=comma, x=x, y expr=\thisrow{T}/1] {data/Couette-01-1e-4-CN-S.csv}; \label{plot:01-CN-S-2}
  
  \addplot[mark=none,dash pattern= on 1pt off 2pt,smooth,color=color3,very thick] table [col sep=comma, x=x, y expr=\thisrow{T}/1] {data/Couette-01-8e-4-USP.csv}; \label{plot:01-USP-1}
  \addplot[mark=none,dash pattern= on 4pt off 4pt,dash phase=4pt,color=color3,thick] table [col sep=comma, x=x, y expr=\thisrow{T}/1] {data/Couette-01-1e-4-USP.csv}; \label{plot:01-USP-2}
  
  \addplot[mark=none,dash pattern= on 1pt off 2pt,smooth,color=color4,very thick] table [col sep=comma, x=x, y expr=\thisrow{T}/1] {data/Couette-01-8e-4-ExpInt.csv}; \label{plot:01-ExpInt-1}
  \addplot[mark=none,dash pattern= on 4pt off 4pt,dash phase=4pt,smooth,color=color4,thick] table [col sep=comma, x=x, y expr=\thisrow{T}/1] {data/Couette-01-1e-4-ExpInt.csv}; \label{plot:01-ExpInt-2}

  \addplot[mark=none,smooth,color=color1,thick] coordinates {(-1, -1) (-2, -2)}; \label{plot:01-1-1}
  \addplot[mark=none,smooth,color=color2,thick] coordinates {(-1, -1) (-2, -2)}; \label{plot:01-2-1}
  \addplot[mark=none,smooth,color=color3,thick] coordinates {(-1, -1) (-2, -2)}; \label{plot:01-3-1}
  \addplot[mark=none,smooth,color=color4,thick] coordinates {(-1, -1) (-2, -2)}; \label{plot:01-4-1}
  \addplot[mark=none,dash pattern= on 4pt off 4pt,smooth,color=black,thick] coordinates {(-1, -1) (-2, -2)}; \label{plot:01-bk-1}
  \addplot[mark=none,dash pattern= on 1pt off 2pt,dash phase=0pt,smooth,color=black,very thick] coordinates {(-1, -1) (-2, -2)}; \label{plot:01-bk-2}

  \coordinate (legend) at (axis description cs:1.00,1.05);
  \end{axis}




  \matrix [
    draw,
    matrix of nodes,
    anchor=south east,
    font=\footnotesize,
    column 1/.style={anchor=base east},
    column 2/.style={anchor=base west},
    column 4/.style={anchor=base east},
    column 5/.style={anchor=base west},
    row sep=0.85mm,
    column sep=0.7mm,
    nodes={inner xsep=0.1mm, inner ysep=0.55mm},
  ] at (legend) {
    \ref*{plot:01-bk-1} & $\Delta t=1\cdot10^{-4}\,\text{s}$ & & \ref*{plot:01-1-1} & CN-SPBGK (ESBGK) \\
    \ref*{plot:01-bk-2} & $\Delta t=8\cdot10^{-4}\,\text{s}$ & & \ref*{plot:01-2-1} & CN-SPBGK (Shakhov) \\
    & & \hspace{0mm} & \ref*{plot:01-4-1}   & ED-SPBGK  \\
    & & \hspace{0mm} & \ref*{plot:01-3-1}  & USPBGK    \\
  };
\end{tikzpicture}
	\caption{Temperature profiles for two different time steps in Case 1 ($Kn=0.1$).}
	\label{fig:couette-01-temperature}
\end{figure}

Figure \ref{fig:couette-01-temperature} shows the temperature profiles at a small and a large time step for the rarefied case $Kn = 0.1$.
Here, the small time step, $\Delta t=\SI{1e-4}{\second}$, corresponds to maximum relaxation factors of $\nu_{ES}\Delta t \approx 0.2$ and $\nu_{Shakhov}\Delta t \approx 0.3$ while the large time step, $\Delta t=\SI{8e-4}{\second}$, corresponds to maximum relaxation factors of $\nu_{ES}\Delta t \approx 1.6$ and $\nu_{Shakhov}\Delta t \approx 2.4$.
At the smaller time step, the time discretization error is negligible.
It is evident that methods solving the Shakhov-BGK equation, specifically CN-SPBGK (Shakhov) and USPBGK, converge to a different solution compared to those solving the ES-BGK equation. This was already discussed in~\cite{Pfeiffer2018,park2024evaluation}.

\begin{figure}
	\centering
	\begin{tikzpicture}

  \begin{axis}[
    scaled ticks=false,
    width=7.5cm,
    height=6cm,
    xmode=log,
    xlabel={$\Delta t / \text{s}$},
    xticklabel style={
      /pgf/number format/fixed,
      /pgf/number format/precision=2
    },
    ymode=log,
    ylabel={$L_{2} / \text{K}$},
    xmin=3e-5,
    xmax=1e-3,
    font=\footnotesize,
  ]

    \addplot[mark=x,color=color1,thick] table [col sep=comma, x=time_step, y expr=\thisrow{L2_norm}/1] {data/Couette-01-L2-CN.csv}; \label{plot:01-CN-L2}

    \addplot[mark=x,color=color2,thick] table [col sep=comma, x=time_step, y expr=\thisrow{L2_norm}/1] {data/Couette-01-L2-CN-S.csv}; \label{plot:01-CN-S-L2}

    \addplot[mark=x,color=color3,thick] table [col sep=comma, x=time_step, y expr=\thisrow{L2_norm}/1] {data/Couette-01-L2-USP.csv}; \label{plot:01–USP-L2}

    \addplot[mark=x,color=color4,thick] table [col sep=comma, x=time_step, y expr=\thisrow{L2_norm}/1] {data/Couette-01-L2-ED.csv}; \label{plot:01-ED-L2}

    \addplot[mark=x,color=color5,thick] table [col sep=comma, x=time_step, y expr=\thisrow{L2_norm}/1] {data/Couette-01-L2-SP.csv}; \label{plot:01-SP-L2}

    \addplot[mark=none,color=black,dashed,thick] coordinates {(1e-6, 4e-5) (1e-3, 4e1)};
    \node at (axis cs:3.8e-4,3) {$\mathcal{O}(\Delta t^2)$};

    \addplot[mark=none,color=black,dashed,thick] coordinates {(1e-5, 4e0) (1e-2, 4e3)};
    \node at (axis cs:1e-4,6e1) {$\mathcal{O}(\Delta t)$};

    \coordinate (legend) at (axis description cs:1.00,1.05);
  \end{axis}

  \matrix [
    draw,
    matrix of nodes,
    anchor=south east,
    font=\footnotesize,
    column 1/.style={anchor=base east},
    column 2/.style={anchor=base west},
    column 4/.style={anchor=base east},
    column 5/.style={anchor=base west},
    row sep=0.85mm,
    column sep=0.7mm,
    nodes={inner xsep=0.1mm, inner ysep=0.55mm},
  ] at (legend) {
    \ref*{plot:01-CN-L2}   & CN-SPBGK (ESBGK)   & & \ref*{plot:01–USP-L2} & USPBGK\\
    \ref*{plot:01-CN-S-L2} & CN-SPBGK (Shakhov) & & \ref*{plot:01-ED-L2}  & ED-SPBGK \\
                          &                    & & \ref*{plot:01-SP-L2}  & SPBGK\\
  };
\end{tikzpicture}
	\caption{Convergence of the $L_2$ error in the temperature for the three different methods in Case 1 ($Kn=0.1$).}
	\label{fig:couette-01-l2}
\end{figure}
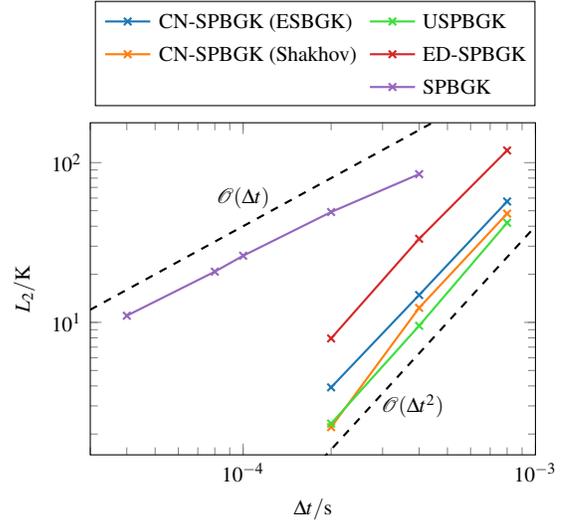

A comparison of the absolute errors at the larger time step reveals that the second-order methods outperform the first-order SPBGK method.
The convergence behavior of the methods can be investigated by considering the $L_2$-error of the temperature, which is defined by
\begin{equation}
L_2 = \sqrt{\sum_{i}{(T_i-T_{\text{exact},i})^2}}.
\end{equation}
Here, the subscript $i$ denotes the values in the corresponding cell.
As the methods converge to different solutions, the "exact" solution, denoted by $T_{\text{exact},i}$, is obtained from simulations utilizing very small time steps for each method.
Figure \ref{fig:couette-01-l2} displays the convergence plots, confirming the expected first-order convergence of the standard SPBGK method.
All other methods, including CN-SPBGK, USPBGK, and ED-SPBGK, demonstrate second-order convergence.

While the ED-SPBGK method exhibits the largest absolute error at a given time step, the Shakhov-based CN-SPBGK and USPBGK methods achieve comparable levels of accuracy.
As noted earlier, the USPBGK method was used without the parameter that switches to standard SPBGK at high Knudsen numbers.
This would otherwise have resulted in it yielding comparable results to SPBGK within this regime.

\subsubsection{Continuum case ($Kn=0.001$)}
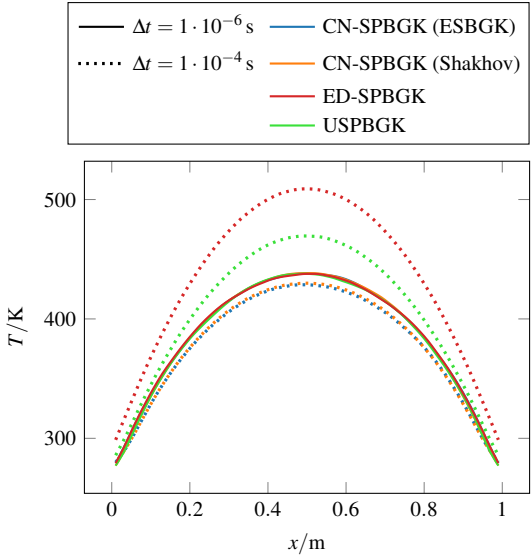
\begin{figure}
	\centering
	\begin{tikzpicture}
  \begin{axis}[
  scaled ticks=false,
  width=7.5cm,
  height=6cm,
  xlabel={$x / \text{m}$},
  xticklabel style={
    /pgf/number format/fixed,
    /pgf/number format/precision=2
  },
  ylabel={$T / \text{K}$},
  xmin=-0.07,
  xmax=1.07,
  cycle list name=exotic,
  font=\footnotesize,
  ]

  \addplot[mark=none,dash pattern= on 1pt off 2pt,dash phase=0pt,smooth,color=color1,very thick] table [col sep=comma, x=x, y expr=\thisrow{T}/1] {data/Couette-0001-1e-4-CN.csv}; \label{plot:0001-CN-1}
  \addplot[mark=none,smooth,color=color1,thick] table [col sep=comma, x=x, y expr=\thisrow{T}/1] {data/Couette-0001-1e-6-CN.csv}; \label{plot:0001-CN-2}

  \addplot[mark=none,dash pattern= on 1pt off 2pt,dash phase=1.5pt,smooth,color=color2,very thick] table [col sep=comma, x=x, y expr=\thisrow{T}/1] {data/Couette-0001-1e-4-CN-S.csv}; \label{plot:0001-CN-S-1}
  \addplot[mark=none,smooth,color=color2,thick] table [col sep=comma, x=x, y expr=\thisrow{T}/1] {data/Couette-0001-1e-6-CN-S.csv}; \label{plot:0001-CN-S-2}
  
  \addplot[mark=none,dash pattern= on 1pt off 2pt,dash phase=0pt,smooth,color=color3,very thick] table [col sep=comma, x=x, y expr=\thisrow{T}/1] {data/Couette-0001-1e-4-USP.csv}; \label{plot:0001-USP-1}
  \addplot[mark=none,smooth,color=color3,thick] table [col sep=comma, x=x, y expr=\thisrow{T}/1] {data/Couette-0001-1e-6-USP.csv}; \label{plot:0001-USP-2}
  
  \addplot[mark=none,dash pattern= on 1pt off 2pt,dash phase=0pt,smooth,color=color4,very thick] table [col sep=comma, x=x, y expr=\thisrow{T}/1] {data/Couette-0001-1e-4-ExpInt.csv}; \label{plot:0001-ED-1}
  \addplot[mark=none,smooth,color=color4,thick] table [col sep=comma, x=x, y expr=\thisrow{T}/1] {data/Couette-0001-1e-6-ExpInt.csv}; \label{plot:0001-ED-2}

  \addplot[mark=none,smooth,color=black,thick] coordinates {(-1, -1) (-2, -2)}; \label{plot:0001-bk-1}

  \coordinate (legend) at (axis description cs:1.00,1.05);
  \end{axis}


  \matrix [
    draw,
    matrix of nodes,
    anchor=south east,
    font=\footnotesize,
    column 1/.style={anchor=base east},
    column 2/.style={anchor=base west},
    column 4/.style={anchor=base east},
    column 5/.style={anchor=base west},
    row sep=0.85mm,
    column sep=0.7mm,
    nodes={inner xsep=0.1mm, inner ysep=0.55mm},
  ] at (legend) {
    \ref*{plot:0001-bk-1} & $\Delta t=1\cdot10^{-6}\,\text{s}$ & & \ref*{plot:0001-CN-2} & CN-SPBGK (ESBGK) \\
    \ref*{plot:01-bk-2} & $\Delta t=1\cdot10^{-4}\,\text{s}$ & & \ref*{plot:0001-CN-S-2} & CN-SPBGK (Shakhov) \\
    & & \hspace{0mm} & \ref*{plot:0001-ED-2}   & ED-SPBGK  \\
    & & \hspace{0mm} & \ref*{plot:0001-USP-2}  & USPBGK    \\
  };

\end{tikzpicture}
	\caption{Temperature profiles for two different time steps in Case 2 ($Kn=0.001$).}
	\label{fig:couette-0001-temperature}
\end{figure}

In the continuum case $Kn = 0.001$, the temperature profiles at two time steps are presented in Figure \ref{fig:couette-0001-temperature}.
In this case, the small time step, $\Delta t=\SI{1e-6}{\second}$, corresponds to maximum relaxation factors of $\nu_{ES}\Delta t \approx 0.2$ and $\nu_{Shakhov}\Delta t \approx 0.3$ while the large time step, $\Delta t=\SI{1e-4}{\second}$, corresponds to maximum relaxation factors of $\nu_{ES}\Delta t \approx 22$ and $\nu_{Shakhov}\Delta t \approx 33$.
In contrast to the rarefied case, all methods converge to a solution that is nearly identical.
This is to be expected, given that the exact solutions of all underlying BGK equations converge to the Navier-Stokes solution as the Knudsen number decreases.

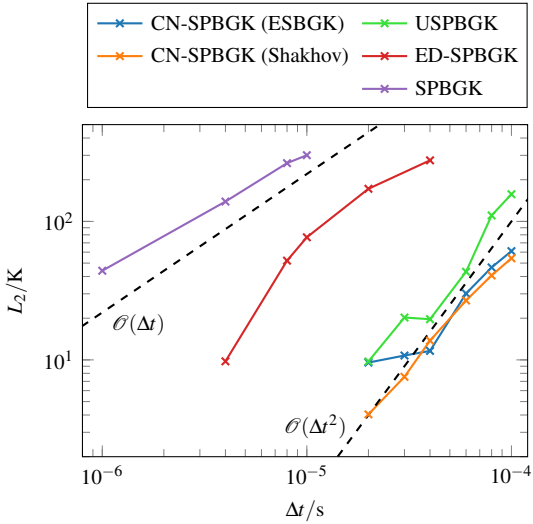
\begin{figure}
	\centering
	\begin{tikzpicture}
  \begin{axis}[
    scaled ticks=false,
    width=7.5cm,
    height=6cm,
    xmode=log,
    xlabel={$\Delta t / \text{s}$},
    xticklabel style={
      /pgf/number format/fixed,
      /pgf/number format/precision=2
    },
    ymode=log,
    ylabel={$L_{2} / \text{K}$},
    ymin=2,
    ymax=5e2,
    xmin=8e-7,
    xmax=1.2e-4,
    font=\footnotesize,
  ]

    \addplot[mark=x,color=color1,thick] table [col sep=comma, x=time_step, y expr=\thisrow{L2_norm}/1] {data/Couette-0001-L2-CN.csv}; \label{plot:0001-CN-L2}

    \addplot[mark=x,color=color2,thick] table [col sep=comma, x=time_step, y expr=\thisrow{L2_norm}/1] {data/Couette-0001-L2-CN-S.csv}; \label{plot:0001-CN-S-L2}

    \addplot[mark=x,color=color3,thick] table [col sep=comma, x=time_step, y expr=\thisrow{L2_norm}/1] {data/Couette-0001-L2-USP.csv}; \label{plot:0001-USP-L2}

    \addplot[mark=x,color=color4,thick] table [col sep=comma, x=time_step, y expr=\thisrow{L2_norm}/1] {data/Couette-0001-L2-ED.csv}; \label{plot:0001-ED-L2}

    \addplot[mark=x,color=color5,thick] table [col sep=comma, x=time_step, y expr=\thisrow{L2_norm}/1] {data/Couette-0001-L2-SP.csv}; \label{plot:0001-SP-L2}

    \addplot[mark=none,color=black,dashed,thick] coordinates {(1e-7, 22e-1) (1e-3, 22e3)};    
    \node at (axis cs:1.5e-6,18) {$\mathcal{O}(\Delta t)$};

    \addplot[mark=none,color=black,dashed,thick] coordinates {(1e-6, 1e-2) (1e-3, 1e4)};
    \node at (axis cs:1.1e-5,3.3) {$\mathcal{O}(\Delta t^2)$};

    \coordinate (legend) at (axis description cs:1.00,1.05);
  \end{axis}

    \matrix [
    draw,
    matrix of nodes,
    anchor=south east,
    font=\footnotesize,
    column 1/.style={anchor=base east},
    column 2/.style={anchor=base west},
    column 4/.style={anchor=base east},
    column 5/.style={anchor=base west},
    row sep=0.85mm,
    column sep=0.7mm,
    nodes={inner xsep=0.1mm, inner ysep=0.55mm},
  ] at (legend) {
    \ref*{plot:0001-CN-L2}   & CN-SPBGK (ESBGK)   & & \ref*{plot:0001-USP-L2} & USPBGK\\
    \ref*{plot:0001-CN-S-L2} & CN-SPBGK (Shakhov) & & \ref*{plot:0001-ED-L2}  & ED-SPBGK \\
                            &                    & & \ref*{plot:0001-SP-L2}  & SPBGK\\
  };
\end{tikzpicture}
	\caption{Convergence of the $L_2$ error in the temperature for the three different methods in Case 2 ($Kn=0.001$).}
	\label{fig:couette-0001-l2}
\end{figure}

The convergence plot in Figure \ref{fig:couette-0001-l2} highlights the performance differences among the methods.
Once more, the standard SPBGK method performs worst and converges with its theoretical first-order in time.
CN-SPBGK and USPBGK maintain their second-order slopes, with CN-SPBGK showing slightly lower $L_2$-errors at equal time steps.
The ED-SPBGK method, although second-order accurate for small time steps, exhibits first-order convergence as the time step size increases, highlighting limitations in its performance for larger time steps due to the increasing influence of the flux term.

\subsection{Hypersonic flow over cylinder} 
\begin{figure}
\centering
\begin{tikzpicture}
\tikzset{
every pin/.style={draw=white,font=\footnotesize, thick, fill opacity=0.45,text opacity=1, text=black},
every pin edge/.style={draw=white, thick},
small dot/.style={fill=white,circle,scale=0.25},
}
\begin{axis}[enlargelimits=false, 
axis on top, 
axis equal image,
scaled x ticks=false, 
xlabel={$x /$ cm},
xticklabel = {
    \pgfmathparse{\tick*100}
    \pgfmathprintnumber{\pgfmathresult}
},
scaled y ticks=false, 
ylabel={$y /$ cm},
yticklabel = {
    \pgfmathparse{\tick*100}
    \pgfmathprintnumber{\pgfmathresult}
},
xmin=0,xmax=0.03,ymin=0,ymax=0.0205,
colormap={rgb}{rgb255=(71,71,219) rgb255=(0,0,91) rgb255=(0,255,255) rgb255=(0,127,0) rgb255=(255,255,0) rgb255=(255,96,0) rgb255=(107,0,0) rgb255=(224,76,76)},
colorbar right,
point meta min=13,
point meta max=2100,
colorbar style={
  font=\footnotesize,
  title={$T /$ K},
  title style={yshift=-2pt,xshift=4pt},
},
font=\footnotesize,
width=0.42\textwidth,
legend style={
	font=\footnotesize,
	legend cell align=left,
	at={(0.95,0.98)},
	anchor=north east,
			fill=none,
			draw=none}]
\addplot graphics [xmin=0,xmax=0.03,ymin=0,ymax=0.0205] {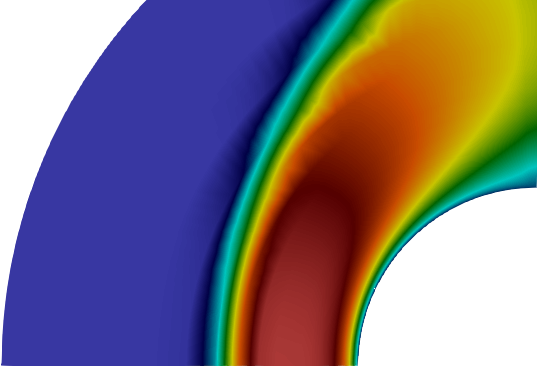};

\end{axis}

\end{tikzpicture}
\caption{Temperature plot of the hypersonic argon flow over a cylinder.\label{fig:temp}}
\end{figure}

A hypersonic flow over a cylinder with radius 1~cm is used with argon to further test the different methods. The inflow conditions are summarized in Table~\ref{tab:cyl}. The temperature of the steady-state solution with CN-ESBGK is depicted in fig.~\ref{fig:temp}.
\begin{table}
\caption{Inflow condition of the hypersonic cylinder case.}\label{tab:cyl}
  \begin{ruledtabular}\renewcommand*{\arraystretch}{1.4}
    \begin{tabular}{ccc}
      $n_\infty \,/\, \SI{}{\per\meter\cubed}$ & $u_{x,\infty} \,/\, \SI{}{\meter\per\second}$ & $T_{\infty} \,/\, \SI{}{\kelvin}$ \\
    $\SI{3.715E21}{}$ & $1502.57$ & $13.3$ \\
    \end{tabular}
  \end{ruledtabular}
\end{table}

Of particular interest for the investigation of the time integration methods is the heat flux on the surface of the cone. On the one hand, the heat flux represents a higher moment and its crucial to produce the correct Prandtl number. On the other hand, it is also relevant from an application perspective for evaluating the efficiency of the methods in such cases. 
As a reference, a DSMC simulation is used, in which both the mean free path and the collision frequency are resolved. This results in a time step of $\Delta t = 5\cdot10^{-9}\,\mathrm{s}$ in the DSMC simulation. In the plots shown in \cref{fig:heatflux} and \cref{fig:heatfluxzoom}, the heat flux along the cylinder surface length $S$ for the different methods are plotted. The smallest time step used is $\Delta t = 1\cdot10^{-7}\,\mathrm{s}$ as CN and USP have already converged at this time step size. 
This corresponds to the largest value of $\nu_{ES}\Delta t_{1E-7}\approx1.2$ and $\nu_{Shakhov}\Delta t_{1E-7}\approx1.8$ in the flow field at the front of the cylinder.
For the Exponential Differencing smaller and for the SPBGK method much smaller time steps are necessary to reach a converged result. It can be observed, both in the plot over the entire surface of the cylinder (see \cref{fig:heatflux}) and in the zoomed-in plot (see \cref{fig:heatfluxzoom}), that CN shows significantly better results than standard SP and the Exponential Differencing method. For very large time steps, CN also demonstrates better results than USP as already seen in the continuum Couette flow, although the performance of both methods is similar for this test case. The biggest used time step corresponds to the largest value of $\nu_{ES}\Delta t_{16E-7}\approx18$ and $\nu_{Shakhov}\Delta t_{16E-7}\approx27$ in the flow field at the front of the cylinder. For this test case, there is basically no difference between the ESBGK and the Shakhov-BGK implementation of the CN method.

\begin{figure}
\centering
\begin{tikzpicture}
\begin{axis}[
scaled ticks=false,
width=7.5cm,
height=6cm,
xlabel={$S / \mathrm{m}$},
ylabel={Heat flux $/ \si{\kilo\watt\per\square\meter}$},
ymin=0,
xmin=0,
legend style={
	font=\footnotesize,
     legend columns = 3,
	legend cell align=left,
	at={(1.,1.1)},
	anchor=south east,
			fill=none,
			draw=black}
]

\addplot[mark=none, color=black] table [col sep=comma, x="arc_length", y expr=\thisrow{"Total_HeatFlux"}/1000] {data/DSMC.csv};
 \addlegendentry{DSMC}
 \addplot[mark=none, color=color5] table [col sep=comma, x="arc_length", y expr=\thisrow{"Total_HeatFlux"}/1000] {data/SP_1E-7.csv};
 \addlegendentry{SPBGK}

\addplot[mark=none, color=color1] table [col sep=comma, x="arc_length", y expr=\thisrow{"Total_HeatFlux"}/1000] {data/CN_1E-7.csv};
 \addlegendentry{CN-SPBGK (ESBGK)}
\addplot[mark=none, color=color3] table [col sep=comma, x="arc_length", y expr=\thisrow{"Total_HeatFlux"}/1000] {data/USP_1E-7.csv};
 \addlegendentry{USPBGK}

  \addplot[mark=none, color=color4] table [col sep=comma, x="arc_length", y expr=\thisrow{"Total_HeatFlux"}/1000] {data/EXP_1E-7.csv};
\addlegendentry{ED-SPBGK}
 \addplot[mark=none, color=color2] table [col sep=comma, x="arc_length", y expr=\thisrow{"Total_HeatFlux"}/1000] {data/CNS_1E-7.csv};
 \addlegendentry{CN-SPBGK (Shakhov)}
 
 \addlegendimage{black}
 \addlegendentry{$1\cdot10^{-7}\,\text{s}$}

 \addlegendimage{dashed,black,thick}
 \addlegendentry{$8\cdot10^{-7}\,\text{s}$}
\addlegendimage{dotted,black,very thick}
 \addlegendentry{$16\cdot10^{-7}\,\text{s}$}
 
\addplot[dashed,dash pattern= on 3pt off 3pt,dash phase=1pt,color=color1,thick] table [col sep=comma, x="arc_length", y expr=\thisrow{"Total_HeatFlux"}/1000] {data/CN_8E-7.csv};
\addplot[dotted, color=color1,very thick] table [col sep=comma, x="arc_length", y expr=\thisrow{"Total_HeatFlux"}/1000] {data/CN_16E-7.csv};

\addplot[dashed,dash pattern= on 3pt off 3pt,dash phase=2pt,color=color3,thick] table [col sep=comma, x="arc_length", y expr=\thisrow{"Total_HeatFlux"}/1000] {data/USP_8E-7.csv};
\addplot[dotted, color=color3,very thick] table [col sep=comma, x="arc_length", y expr=\thisrow{"Total_HeatFlux"}/1000] {data/USP_16E-7.csv};

\addplot[dashed, color=color5,thick] table [col sep=comma, x="arc_length", y expr=\thisrow{"Total_HeatFlux"}/1000] {data/SP_8E-7.csv};

\addplot[dashed, color=color4,thick] table [col sep=comma, x="arc_length", y expr=\thisrow{"Total_HeatFlux"}/1000] {data/EXP_8E-7.csv};

\addplot[dashed, color=color2,thick] table [col sep=comma, x="arc_length", y expr=\thisrow{"Total_HeatFlux"}/1000] {data/CNS_8E-7.csv};
\addplot[dotted, color=color2,very thick] table [col sep=comma, x="arc_length", y expr=\thisrow{"Total_HeatFlux"}/1000] {data/CNS_16E-7.csv};



\end{axis}
\end{tikzpicture}
\caption{Heat flux profiles on the cylinder surface for 3 different time steps.\label{fig:heatflux}}
\end{figure}
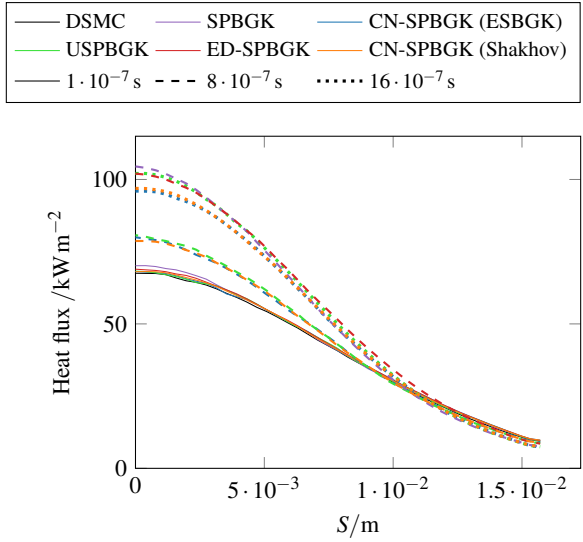
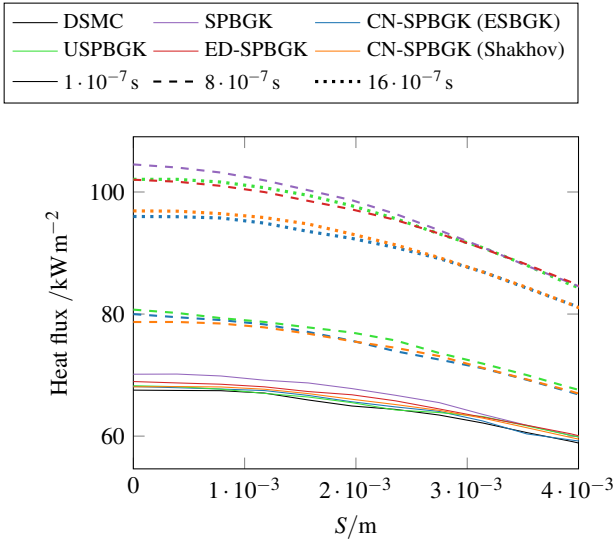
\begin{figure}
\centering
\begin{tikzpicture}
\begin{axis}[
scaled ticks=false,
width=7.5cm,
height=6cm,
xlabel={$S / \mathrm{m}$},
ylabel={Heat flux $/ \si{\kilo\watt\per\square\meter}$},
xmin=0,
xmax=4E-3,
legend style={
	font=\footnotesize,
     legend columns = 3,
	legend cell align=left,
	at={(1.,1.1)},
	anchor=south east,
			fill=none,
			draw=black}
]

\addplot[mark=none, color=black] table [col sep=comma, x="arc_length", y expr=\thisrow{"Total_HeatFlux"}/1000] {data/DSMC.csv};
 \addlegendentry{DSMC}
\addplot[mark=none, color=color5] table [col sep=comma, x="arc_length", y expr=\thisrow{"Total_HeatFlux"}/1000] {data/SP_1E-7.csv};
 \addlegendentry{SPBGK}
 
\addplot[mark=none, color=color1] table [col sep=comma, x="arc_length", y expr=\thisrow{"Total_HeatFlux"}/1000] {data/CN_1E-7.csv};
 \addlegendentry{CN-SPBGK (ESBGK)}
\addplot[mark=none, color=color3] table [col sep=comma, x="arc_length", y expr=\thisrow{"Total_HeatFlux"}/1000] {data/USP_1E-7.csv};
 \addlegendentry{USPBGK}

  \addplot[mark=none, color=color4] table [col sep=comma, x="arc_length", y expr=\thisrow{"Total_HeatFlux"}/1000] {data/EXP_1E-7.csv};
\addlegendentry{ED-SPBGK}
 \addplot[mark=none, color=color2] table [col sep=comma, x="arc_length", y expr=\thisrow{"Total_HeatFlux"}/1000] {data/CNS_1E-7.csv};
 \addlegendentry{CN-SPBGK (Shakhov)}
 
 \addlegendimage{black}
 \addlegendentry{$1\cdot10^{-7}\,\text{s}$}

 \addlegendimage{dashed,black,thick}
 \addlegendentry{$8\cdot10^{-7}\,\text{s}$}
\addlegendimage{dotted,black,very thick}
 \addlegendentry{$16\cdot10^{-7}\,\text{s}$}
 
\addplot[dashed, color=color1,thick] table [col sep=comma, x="arc_length", y expr=\thisrow{"Total_HeatFlux"}/1000] {data/CN_8E-7.csv};
\addplot[dotted, color=color1,very thick] table [col sep=comma, x="arc_length", y expr=\thisrow{"Total_HeatFlux"}/1000] {data/CN_16E-7.csv};

\addplot[dashed, color=color3,thick] table [col sep=comma, x="arc_length", y expr=\thisrow{"Total_HeatFlux"}/1000] {data/USP_8E-7.csv};
\addplot[dotted, color=color3,very thick] table [col sep=comma, x="arc_length", y expr=\thisrow{"Total_HeatFlux"}/1000] {data/USP_16E-7.csv};

\addplot[dashed, color=color5,thick] table [col sep=comma, x="arc_length", y expr=\thisrow{"Total_HeatFlux"}/1000] {data/SP_8E-7.csv};

\addplot[dashed, color=color4,thick] table [col sep=comma, x="arc_length", y expr=\thisrow{"Total_HeatFlux"}/1000] {data/EXP_8E-7.csv};

\addplot[dashed, color=color2,thick] table [col sep=comma, x="arc_length", y expr=\thisrow{"Total_HeatFlux"}/1000] {data/CNS_8E-7.csv};
\addplot[dotted, color=color2,very thick] table [col sep=comma, x="arc_length", y expr=\thisrow{"Total_HeatFlux"}/1000] {data/CNS_16E-7.csv};



\end{axis}
\end{tikzpicture}
\caption{Detail of the heat flux profiles close to the stagnation point for 3 different time steps.\label{fig:heatfluxzoom}}
\end{figure}

\section{Conclusion}
Particle methods represent an appealing class of methods widely used for simulating rarefied gas dynamics. Despite their robustness and popularity, these methods typically exhibit numerical accuracy and convergence limited to first order. However, accurately capturing multi-scale flow phenomena in realistic rarefaction regimes demands high-order schemes, necessitating significant improvements in the numerical framework of stochastic particle methods.
This study introduces advancements to SPBGK schemes by implementing a consistent second-order treatment of the BGK relaxation using the Crank-Nicolson scheme with an Grad-13 approximation for large time steps. The resulting CN-SPBGK method for the ESBGK and Shakhov-BGK models is designed to be straightforward to implement, with minimal computational overhead compared to traditional SPBGK approaches. The Grad-13 approximation of the CN-SPBGK operator also allows large time steps without the introduction of negative weights. 
The presented CN-SPBGK method demonstrates significantly better results compared to the second-order Exponential Differencing method, primarily due to its superior approximation of the flux term, which naturally emerges from the CN approach without requiring additional interventions. For small Knudsen numbers, the method also shows slightly better performance than the effective USP method for large time steps. 

An additional advantage is that no extra parameter related to the local Knudsen number is required in the CN method, and its derivation is somewhat more straightforward compared to the USP method. Specifically, no additional collision term is needed in the particle motion during the derivation process. Combined with the introduced Grad-13 approximation of the collision term, this facilitates a straightforward extension of the particle-based CN method to BGK models with internal excitation energies for molecules or multi-species models. 

Moreover, coupling with other methods, as in \cite{PhysRevE.108.015302}, is also feasible without imposing time-step limitations on the particles to avoid generating negative pre-factors and therefore negative particle weights.

\section*{Acknowledgments}
  This project has received funding from the European Research Council (ERC) under the European Union's Horizon 2020 research and innovation programme (grant agreement No. 899981 MEDUSA).

\appendix 
\section{Flux of the CN-SPBGK method}
\label{sec:appflux}
As described in \citet{fei2020unified}, the average numerical flux at position $\mathbf{x}$ over a time step $\Delta t$ is given as
\begin{equation}
    \mathbf{F}=\frac{1}{\Delta t}\int_0^{\Delta t} \mathbf{v} f(\mathbf{x},\mathbf{v},t)\,dt.
\end{equation}
In the CN-SPBGK method, due to the leapfrog behavior it can be written
\begin{equation}
    \mathbf{F}=\frac{1}{\Delta t}\int_0^{\Delta t} \mathbf{v}\left( f(\mathbf{x}-\mathbf{v}t,\mathbf{v},0)+\frac{\Delta t}{2} \Omega(\mathbf{x}-\mathbf{v}t,\mathbf{v},0)\right)\,dt.
\end{equation}

We now expand around $\mathbf{x}$, with $s$ as the time integration variable for clarity:

\begin{equation}
\begin{split}
    \mathbf{F}&=\!\begin{multlined}[t][8cm]
        \frac{1}{\Delta t}\int_0^{\Delta t} \mathbf{v}\left( f(0)-s\left(\mathbf{v}\frac{\partial f}{\partial\mathbf{x}}(0)\right)\right.\\
        \left. +\frac{\Delta t}{2} \left(\Omega(0)-s\mathbf{v}\frac{\partial \Omega(0)}{\partial\mathbf{x}}\right)+\mathcal{O}(s^2)\right)\,ds\qquad
    \end{multlined}\\
    &= \!\begin{multlined}[t]
        \frac{1}{\Delta t}\int_0^{\Delta t} \mathbf{v}\bigg( f(0)+s\left(\frac{\partial f}{\partial t}(0) -\frac{1}{\tau}(f^t(0)-f(0))\right)\\
        +\frac{\Delta t}{2} \Omega(0)+\mathcal O(s\Delta t)+\mathcal{O}(s^2)\bigg)\,ds
    \end{multlined}
\end{split}
\label{eq:fluxtime}
\end{equation}

with $f(0)= f(\mathbf{x},\mathbf{v},0)$.

To obtain the asymptotic behavior of this flux in the continuum limit, following the Chapman-Enskog proof for the ESBGK model given by \citet{andries2000gaussian}, we rearrange the BGK equation and use a first-order expansion of $f$, giving:
\begin{equation}
    f = f^t - \tau D_t f^t + \mathcal{O}(\tau^2)\quad \textrm{with}\quad D_t = \frac{\partial}{\partial t} + \mathbf{v}\cdot \frac{\partial}{\partial \mathbf x}.
\end{equation}
Using this in Eq.~\eqref{eq:fluxtime} yields:
\begin{equation}
\begin{split}
    \mathbf{F}&=\!\begin{multlined}[t][8cm]
    \frac{1}{\Delta t}\int_0^{\Delta t} \mathbf{v}\bigg(f^t(0)+\tau D_t f^t(0)+s \frac{\partial f^t}{\partial t}(0)\\
    -\frac{s}{\tau}\left(\tau D_t f^t(0)\right) + \frac{\Delta t}{2\tau}\left(\tau D_t f^t(0)\right)\\
    + \mathcal O(\tau^2) + \mathcal O(s \tau)+ \mathcal O(s\Delta t)+\mathcal O(s^2)\bigg)\,ds.
    \end{multlined}
\end{split}    
\end{equation}

The integration results in
\begin{equation}
    \mathbf{F}=\!\begin{multlined}[t]\mathbf{v}\bigg(f^t(0)+\frac{\Delta t}{2}\frac{\partial f^t}{\partial t}(0)-\tau D_t f^t(0)\\
    +\mathcal{O}(\tau^2)+\mathcal{O}(\tau \Delta t)+\mathcal{O}(\Delta t^2)\bigg)
    \end{multlined}
\end{equation}
which corresponds to the Navier-Stokes flux up to the $\mathcal{O}(\Delta t^2)$ term~\cite{fei2020unified,struchtrup2005macroscopic}.

\bibliography{mybibfile}

\end{document}